%% file: paper.tex
\documentclass[useAMS,usenatbib]{mn2e}
\pdfoutput=1
\usepackage{amsmath}
\usepackage{amssymb}
\usepackage{graphicx}
\usepackage[usenames,dvipsnames]{color}

\usepackage{hyperref}

\newcommand{\HI}{{\text{H\MakeUppercase{\romannumeral 1}}} }
\newcommand{\HII}{{\text{H\MakeUppercase{\romannumeral 2}}} }
\newcommand{\Lya}{Ly$\alpha$ }
\newcommand{\ccm}{\,\mathrm{cm}^{-3}}
\newcommand{\kms}{\,\mathrm{km}\,\mathrm{s}^{-1}}
\input{ads_commands}

\title[Directional \Lya Equivalent Boosting]{Directional \Lya equivalent boosting I. Spherically symmetric distributions of clumps}
\author[M. Gronke and M. Dijkstra]{M. Gronke$^{1}$\thanks{E-mail:
maxbg@astro.uio.no} and M. Dijkstra$^{1}$\\
$^{1}$Institute of Theoretical Astrophysics, University of Oslo, Postboks 1029, 0315 Oslo, Norway}

\begin{document}

\date{Accepted 2014 July 25. Received 2014 July 9; in original form 2014 June 12}

\pagerange{\pageref{firstpage}--\pageref{lastpage}} \pubyear{2014}

\maketitle

\label{firstpage}

\begin{abstract}
We quantify the directional dependence of the escape fraction of Lyman-$\alpha$ (Ly$\alpha$) and non-ionizing UV-continuum photons from a multiphase medium, and investigate whether there exist directional enhancements in the Ly$\alpha$ equivalent width (EW). Our multiphase medium consists of spherically symmetric distributions of cold, dusty clumps embedded within a hot dust-free medium. We focus on three models from the analysis presented by Laursen et al. (2013). We find that for a Ly$\alpha$ and UV-continuum point source, it is possible to find an EW boost $b(\theta,\phi) > 5 \bar{b}$ in a few per cent of sight lines, where $\bar{b}$ denotes the boost averaged over all photons. For spatially extended sources this directional dependence vanishes quickly when the size of the UV emitting region exceeds the mean distance between cold dusty clumps.  Our analysis suggests that directional EW boosting can occur, and that this is mostly driven by reduced escape fractions of UV photons (which gives rise to UV-continuum `shadows'), and less due to an enhanced Ly$\alpha$ escape fraction (or beaming),  in certain directions.
\end{abstract}

\begin{keywords}
radiative transfer -- ISM: clouds -- galaxies: ISM -- line: formation -- scattering -- galaxies: high-redshift
\end{keywords}

\section{Introduction}
Lyman-$\alpha$ emitting galaxies are an important probe of the high-redshift Universe. Their strong \Lya radiation makes them easy to find in narrowband surveys, and allows for easier spectroscopic follow-up in order to confirm their redshift. The most distant Lyman-$\alpha$ emitting galaxies have been found out to $z\sim 7.6$ \citep{2006Natur.443..186I, 2013Natur.502..524F, 2014arXiv1404.4632S, Ono2012}.

One of the fundamental quantities associated with the Ly$\alpha$ line is its equivalent width (EW), which is defined as the ratio of the total line flux and the surrounding flux density. \Lya radiation from galaxies is predominantly emitted as hydrogen recombination radiation inside \HII regions surrounding young, hot stars \citep{1967ApJ...147..868P}. Synthetic stellar models can therefore be used to predict its strength \citep{2003A&A...397..527S,2010A&A...523A..64R}. These models predict that `conventional' stellar populations can produce a maximum EW$_{\rm max}=240\,$\AA\,\,\citep[][and references therein]{Laursen2012}. Larger EW values have been observed \citep[e.g.][]{Dawson2004,Nilsson2010,Kashikawa2012}, which has led to speculation that these galaxies may harbour `unusual' (extremely metal-poor or even metal-free) stellar populations \citep{2002ApJ...565L..71M,2007MNRAS.379.1589D}.

However, radiative transfer of Ly$\alpha$ photons through a multiphase medium is a complex process which can affect the EW in non-trivial ways. \citet{Neufeld1991} argued that multiphase gas, consisting of cold dusty clumps embedded with a hot tenuous medium, can facilitate the escape of Ly$\alpha$ photons from dusty media, and `under special conditions' boost the EW. This boost requires an enhanced chance of escape for \Lya photons compared to (non-ionizing) UV-continuum photons. The so-called `Neufeld scenario/mechanism' relies on the fact that hydrogen atoms can only (efficiently) scatter Ly$\alpha$ photons, which can prevent \Lya from penetrating \HI clouds. If dust resides in these clouds, \Lya photons can effectively avoid being absorbed by dust grains. Consequently, \Lya photons propagate mainly through the inter-cloud medium (ICM) which contains little or no dust. The UV-continuum radiation, on the other hand, is not affected by the `shielding' and instead penetrates deep into the clouds where the chance of destruction is much higher. 

\citet{Hansen2005} explored the Neufeld-mechanism numerically with a \Lya radiative transfer code, and showed that quantitatively dusty clumps can lead to EW-boosting. Recent studies have explored under what physical conditions EW boosting occurs: \citet{Duval2013} found several requirements had to be fulfilled for activating the Neufeld-mechanism: a practically empty ICM, slow \& uniform galactic outflows, high dust density within the clumps and a large number/covering factor of the clumps, and concluded that such conditions are ``quite unlikely/difficult to find''.  \citet{Laursen2012} (hereafter, L13) obtained a similar conclusion after studying a wide range of more realistic models. L13 conclude that they consider ``the Neufeld model to be an extremely unlikely reason for the observed high EWs''. Although L13 comment on directional variations in the EW-boost, $b$ has so far always calculated as an average value, denoted with $\bar{b}$. This simplification is understandable, as resolving the directional dependence requires a much larger number of photons in the Monte Carlo (MC) simulation, which makes it practically impossible to explore a large parameter space (which was the goal in \citet{Laursen2012} \& \citet{Duval2013}). However, it is important to recall that observations do not measure an averaged boost parameter per galaxy, but instead probe $b$ in a specific direction. This means even when realistic models predict an average EW boost of $\bar b\sim 1$, there may be distinct lines of sight with larger values of $b$. {\it As a consequence of this directional dependence, radiative-transfer induced EW-boosting may still provide a relevant explanation for large EWs, even when the angle-averaged value suggests that no-boost occurs}.\\

Throughout this paper, we define the directional dependent boost factor $b(\theta,\phi)$ as
\begin{equation}
b(\theta,\phi) \equiv \frac{f_{{\rm esc},Ly\alpha}(\theta,\phi)}{f_{{\rm esc},UV}(\theta,\phi)}, 
\end{equation} where $f_{{\rm esc}, Ly\alpha}(\theta,\phi)$ denotes the escape fraction of \Lya photons in direction $(\theta,\phi)$, $f_{{\rm esc},UV}(\theta,\phi)$ that of UV-continuum photons\footnote{\citet{2008ApJ...678..655F,2009ApJ...691..465F} characterize the EW boost with a `$q$ parameter' which is defined as $f_{{\rm esc},Ly\alpha}\equiv \exp(-q\tau_{\rm a,{\rm eff}})$. Here, $\tau_{a,{\rm eff}}$ is an `effective' absorption optical depth given by $f_{{\rm esc, UV}} \equiv \exp(-\tau_{a,{\rm eff}})$. The parameter $q$ is therefore related to $b$ via $ b = \exp(-q\tau_{\rm a,{\rm eff}})/\exp(-\tau_{\rm a,{\rm eff}})$, and $b>1$ [$b <1$] corresponds to $q<1$ $[q > 1]$.}. We refer to the ``average'' boost, $\bar{b}$, as the boost calculates using \textit{all} photons. This `photon-weighted' boost has been the subject of previous studies. We stress that this average quantity can be different from the `angle-averaged' boost -- defined as $\langle b \rangle \equiv \frac{1}{4\pi} \int d\Omega b(\theta,\phi)$ -- as $\bar{b}$ puts more weight on sight lines along which more photons escape.

In order to study the directional dependence of $b(\theta,\phi)$, we bin the \Lya and UV-continuum escape fractions in equiareal bins using the \texttt{HEALPix}\footnote{http://healpix.sourceforge.net} tessellation and obtain a measure for $b(\theta,\phi)$.  As mentioned previously, this task requires a large amount of photons per geometrical setup, and we therefore focus on only three sets of parameters.

In the current paper, we focus on spherical distributions of clumps following Laursen et al. (2013, and also Duval et al. 2014).  Clearly, anisotropic gas distributions (such as discs, biconical outflows, etc) will give rise to more anisotropic escape of Ly$\alpha$ photons, and we will study this in a forth-coming paper. This study contains the first calculations of Ly$\alpha$ and UV-continuum escape over the full $2$D plane of the sky. 
Previous studies either presented azimuthally averaged quantities \citep{2007ApJ...657L..69L,2012A&A...546A.111V,Zheng2013,Behrens2014} or focused solely on \Lya escape \citep{2012ApJ...754..118Y}. Our study shows that generating 2D maps of EW-boosting introduces new complications to the analysis. The benefit of using spherical clump distributions is that it allows us to ({\it i}) connect easily to previous studies \citep{Laursen2012,Duval2013}, and ({\it ii}) more easily identify key quantities that are relevant for directional EW-boosting.\\

The paper is organized as follows: in Sec.~\ref{sec:method} we give an overview over the model parameters, present our models, and, briefly outline our new radiative transfer code.  We present our results in Sec.~\ref{sec:results} and discuss them in Sec.~\ref{sec:discussion}. Finally, we conclude in Sec.~\ref{sec:conclusion}.

\section{Method}
\label{sec:method}

\subsection{The model}
To facilitate comparison with previous work, we closely follow the analysis of L13.  In their analysis, a galaxy is represented by a spherical distribution of clumps embedded within a hot dust-free medium, both of which extend out to radius $r_{\rm gal}$. L13 provide an extensive analysis of what are reasonable values for the parameters describing the multiphase interstellar medium of this galaxy. Specifically, we adopt the parameters shown in Table~1 and described in section~3 of their paper. These parameters fall in four categories:
\begin{enumerate}
\item Inter-Cloud Medium (ICM) Parameters. Parameters defining the hot ICM are the hydrogen number density $n_{\HI,\mathrm{ICM}}$, and the temperature $T_{\rm ICM}$. The ICM extends out to $r_{\rm gal}$ and is assumed to be dust-free.
\item Emission Parameters. The initial distance from the centre of the simulation box/galaxy is drawn from $r \sim \exp(-r / H_{\rm em})$ distribution. Additionally, photons are emitted with a probability of $P_{\rm cl}$ within a cloud. The emission frequency is drawn from a Gaussian with standard deviation $\sigma_{\rm em}$.
\item Cloud geometry. L13 ran roughly half of the simulations with constant cloud sizes, while in the other half, cloud sizes were drawn from an exponential distribution. In the following, we focus solely on clouds wit a constraint size $r_{\rm cl}$. L13 found this choice had little impact on the radiative transfer. A key parameter is the covering factor $f_{\rm c}$, defined as the average number of clouds from the centre to the edge of the distribution (as in L13)\footnote{This differs from the definition for $f^{\rm DK12}_{\rm c} \equiv n_{\rm cl}\pi r^2_{\rm cl}$ adopted in Dijkstra \& Kramer (2012), in which $n_{\rm cl}$ denotes the number density of clouds. The two covering factors are related via $f_{\rm c} = \int_0^{r_{\rm gal}}{\rm d}s\, f^{\rm DK12}_{\rm c}(s)$.}. The volume filling factor $F_{\rm cl}$ of clumps -- which is the fraction of the galactic volume covered by clouds -- relates to the covering factor as $F_{\rm cl}=4 r_{\rm cl}f_{\rm c}/ (3 r_{\rm gal})$.

This category also contains the parameters for the \HI number density inside the cloud, $n_{\HI, \mathrm{cl}}$, the gas temperature in the clouds, $T_{\rm cl}$, and their dust content. We quantity the dust content of a cloud via its `absorbing' optical depth from the centre to the surface of a cloud, $\tau_{a,{\rm cl}}$. This absorbing optical depth relates to the total optical depth in dust, $\tau_{d,\rm{cl}}$,  via $\tau_{a,{\rm cl}}=(1-A)\tau_{d,{\rm cl}}$, where $A$ is the dust albedo. Throughout this work, we used $A=0.32$ \citep{2001ApJ...554..778L}. The total dust optical depth relates to gas metallicity $Z$ as $\tau_{d,cl} = n_{\HI,cl}r_{cl}\sigma_d Z/Z_\odot$.
Here, $Z_\odot$ and $\sigma_d$ denote the Sun's metallicity and the dust cross-section, respectively.

\item Cloud motion. Cloud motion plays a crucial role in how deep the \Lya photons travel into the clouds and, hence, their chance of absorption. As L13, we allow two types of cloud motion: ({\it a}) a symmetric outflow with the radial dependent velocity. The magnitude of the outflow is
\begin{equation}
  v(r) = v_\infty\left\{1 - \left(\frac{r}{r_{\text{min}}}\right)^{1-\alpha}\right\}^{1/2}
\end{equation}
 for $r>r_{\text{min}}$ and otherwise zero. This three parameter outflow model follows from the assumption that the acceleration of the cloud decreases proportional to $r^{-\alpha}$ \citep[as introduced by][]{Steidel2010}. Instead of using $V_{out}=v(r_{\rm gal})$, we stick to the notation by \citet{Kramer2012} with $v_\infty = v(r\rightarrow\infty)$; ({\it b})  a superimposed velocity dispersion with standard deviation $\sigma_{\rm cl}$.
\end{enumerate}
The twelve variables described above span a vast parameter space which has been explored intensively by L13. Instead of trying to cover as much of the parameter space as possible, here we focus on three ``cornerstone'' parameter sets.

\begin{itemize}
\item First, we study the fiducial model from L13. The fiducial model contains $6500$ clouds with $r_{\rm cl}=100\,$pc out to $r_{\rm gal}=5\,$kpc. This results in a volume filling factor of $F_{\rm cl}=0.052$ and a covering factor $f_{\rm c}=2$. The clouds contain a hydrogen number density $n_{\HI ,\mathrm{cl}} = 1\ccm$, and dust resulting in $\tau_{a,{\rm cl}}=0.48$. This model was designed to be as simple as possible, and we assume that the ICM is empty, that the clouds are static, and that all photons are emitted from a central source with an initial frequency of $x_i = 0$. Furthermore, we set the temperatures to $T_{\rm cl}=10^4$K and $T_{\rm ICM}=10^6$K. The photon-averaged EW boost for the fiducial model is $\bar{b}\approx 2$ (Fig.~2 in L13).

\item We chose the second set of parameters to lie within a range which L13 describes as `realistic'.  This range is selected to represent a model that is closer to reality. This model has $(n_{\HI ,{\rm cl}}, T_{\rm cl}, \tau_{a, {\rm cl}}, F_{\rm cl}, n_{\HI,{\rm ICM}}, T_{\rm ICM}) = (0.35\ccm, 1.4\times 10^4\mathrm{K}, 2.35\times 10^{-2}, 0.104, 10^{-9}\ccm, 10^6\mathrm{K})$, which lies in the centre of the parameter range quoted to be reasonable by L13. The parameters $r_{\rm cl}$ and $r_{\rm gal}$ are unchanged compared to the fiducial model. The main differences are that the clouds have a velocity dispersion of $\sigma_{\rm cl}=40\kms$, a ``momentum based outflow'' with parameters $(v_\infty, r_{\mathrm{min}}, \alpha) = (40\kms, 1\mathrm{kpc}, 1.5)$, the photons are not emitted at the same place but the prior introduced exponential emission site distribution is employed with $H_{\rm em}=1\,$kpc, and roughly a third of the photons were forced to start their propagation within a cloud, i.e., $P_{\rm cl}=0.35$. Moreover, the photon is not emitted with $x_i=0$ but instead the emission line width is $\sigma_{\rm em}=40\kms$. These values lead to a value of $\bar b\lesssim 1$ (see Fig.~15 in L13).

\item The third parameter set lies in the range which L13 label as ``extreme, but possibly conceivable'' and is, therefore, labelled as `extreme model'. The cloud motion mode as well as the emission site distribution is the same as in the `realistic model' but the parameters are altered to $(n_{\HI ,{\rm cl}},\linebreak[0] T_{\rm cl},\linebreak[0] \tau_{a, {\rm cl}},\linebreak[0] F_{\rm cl}, n_{\HI,{\rm ICM}}, T_{\rm ICM}, \sigma_{\rm cl}, v_\infty, \sigma_{\rm em}, P_{\rm cl}) = (1.5\ccm,\linebreak[0] 10^4\,\mathrm{K},\linebreak[0] 0.55,\linebreak[0] 0.08,\linebreak[0] 10^{-10}\ccm,\linebreak[0] 10^6\,\mathrm{K},\linebreak[0] 14\kms,\linebreak[0] 23\kms,\linebreak[0] 15\kms,\linebreak[0] 0.15)$.  L13 showed that models with such a high hydrogen density within the clouds, and a close to empty ICM have a boost parameter of $\bar{b}>1$.
\end{itemize}

We are interested in the value of $b(\theta,\phi)$ for individual sight lines, and the directional binning is of importance. Since the solid angle of a telescope as seen from the \Lya emitting galaxy is basically zero, the number of pixels should be as high as possible. On the other hand, to prevent Poisson-noise from dominating our statistics, we would like the number of photons per directional bin to be $\gtrsim 100$. In order to find the right balance between resolution and computation time, we ran the fiducial model with $\sim 10^9$ \Lya and $\sim 10^{10}$ UV photons and increase the number of bins until the $b$ distribution reaches convergence (when this is possible: we will see below that this is not always the case).

\subsection{Radiative transfer code}
The computations were performed with a newly created Monte Carlo (MC) radiative transfer code \texttt{tlac}. The code was developed after the ingredients described, e.g., in \citet{Laursen2010} and \citet{Dijkstra2014arXiv1406.7292D}.
We adopt the standard convention of expressing frequencies in terms of $x\equiv (\nu - \nu_0)/\Delta\nu_D = c (\nu/\nu_0 - 1)/v_{\rm th}$ -- where $v_{\rm th}$ is the thermal velocity of the hydrogen atoms and $\nu_0$ the \Lya  line centre.

We emit a photon in a random direction $\vec{k}$. We then generate $\tau$ from the distribution $P(\tau)=\exp(-\tau)$. We convert $\tau$ into the distance $d$ the photon propagates before interacting with a hydrogen or dust particle via $\tau =\int_0^d \mathrm{d}s\, [\sigma_{\HI} n_\HI(s) + \sigma_dn_d(s)]$. We decide whether a photon interacts with \HI or dust by comparing a uniform random variable $\mathcal{R}$ to the probability for interaction with dust $P_{\rm dust}=\sigma_d n_d / (\sigma_\HI n_\HI + \sigma_dn_d)$. In case of dust-interaction, we compare a second uniform random variable to the dust albedo $A$ to determinate if a scattering or absorption occurs.

We generate a new random direction of the photon from the proper `phase function' $P(\mu)\mathrm{d}\mu$, which gives the probability that the cosine between the old and the new direction $\mu \equiv {\bf k}_{\rm in} \cdot {\bf k}_{\rm out}$ lies in the range $\mu \pm \mathrm{d}\mu/2$. Here, ${\bf k}_{\rm in}$ (${\bf k}_{\rm out}$) denotes the propagation direction before (after) scattering. For resonant scattering via the $2P_{1/2}$ state (probability of $1/3$) gives a rise to uniform scattering, whereas scattering via the $2P_{3/2}$ state (probability of $2/3$) results in a scattering with $P_{\text{core},3/2}(\mu)=7/16 (1+3\mu^2/7)$.
In the case of wing scattering $P_{\text{wing}}(\mu)=3/8(1+\mu^2)$. Dust scattering is implemented using the \citet*{Henyey1941} phase function with an asymmetry parameter of $g=0.73$. 

The alternation of the photon's direction combined with the movement of the interacting particle relative to the observer results in a change of frequency $x\rightarrow x'$ according to
\begin{equation}
x' = x - u_\parallel + u_\parallel \mu  + u_\perp \sqrt{1 - \mu^2}\;.
\end{equation}
We choose a reference frame so that the velocity of the atom in units of $v_{\rm th}$ is $(u_\parallel, u_\perp, 0)$  and  $u_\parallel$ is aligned to the incoming photon's direction. Furthermore, $u_\parallel$ is generated from
\begin{equation}
  u_\parallel \sim e^{-u_\parallel^2}/\left((x - u_{\parallel})^2 + a^2\right)
\end{equation}
and $u_\perp$ is drawn from a centreed Gaussian with standard deviation of $\sigma=1/\sqrt{2}$. Using a truncated Gaussian instead speeds up the code tremendously \citep{Ahn2001a}. However, care should be exercised when applying this ``core skipping technique'', as it may force photons  unnaturally deep into the clouds which leads to lower values of $b$ (see L13 for details). To be conservative we completely relinquish the acceleration scheme.\\

In order to calculate radiative transfer within a clumpy medium, several algorithms have been developed and employed previously. One approach is to treat the propagation through the ICM exactly and whenever a cloud is encountered, the cloud is treated as a virtual particle with its own phase function and redistribution function. This method was used successfully by \citet{Hansen2005}. In this case photon trajectories are followed only in the hot interclump medium which keeps the computational cost low. The downside of this approach is that the phase-and-redistribution functions depend on the clump properties (e.g. whether there are density and/or velocity gradients), and would need to be evaluated first. L13 took a different approach and modeled the path of the photons within the clouds too, as the code utilized by L13  supports a refined mesh structure. This made it possible to model spherical clouds with a high refinement level at the their surfaces. 
A third possibility for cloud treatment is to handle clouds as virtual particles, similar to the first approach, but to explicitly track photons inside clumps.
Thus, whenever a photon encounters a cloud, the cloud is represented by concentric spherical shells through which we propagate the photon\footnote{This allows us to include velocity and/or density gradients inside clouds, but these were absent in our current analysis.}.
This gives the full solution without the memory requirement needed in order to resolve each cloud separately. This approach was used by \citet{Kramer2012} and will be also employed in this work. 

Our code was tested intensively, and was successfully compared against analytic solutions for the Ly$\alpha$ spectra emerging from extremely optically thick slabs \citep[as in][]{Neufeld1990,Harrington1973} and spheres \citep[as in][]{Dijkstra2006}. It passed the ICM propagation tests described in \citet{Kramer2012}. We also computed $\mathcal{N}_0$ -- the average number of spherical clumps encountered in the absence of dust -- as a function of $f_{\rm c}$ and found that it agreed perfectly with the calculations by \citet{Hansen2005}, who found that $\mathcal{N}_0=f_{\rm c}^2+\frac{4}{5}f_{\rm c}$ (see Eq.~62 in Hansen \& Oh 2006). Finally, we compared the results obtained for the fiducial model to the code developed for the analysis in \citep{Kramer2012}, and found good agreement.

\begin{table}
  \centering
  \caption{Overview of the results from the three models.}
  \begin{tabular}{l|rrrrr}
\hline
     &  $N_{p,Ly\alpha}$ & $N_{p,UV}$ & $\bar f_{{\rm esc},Ly\alpha}$ & $\bar f_{{\rm esc},UV}$ & $\bar b$ \\
\hline\hline
     fiducial & $1.1\times 10^8$ & $1.5\times 10^9$ & $0.93$ & $0.38$ & $2.4$ \\
     realistic & $2.5\times 10^7$ & $6.8\times 10^8$ & $0.74$ & $0.89$ & $0.84$ \\
     extreme & $3.0\times 10^7$ & $4.0\times 10^9$ & $0.33$ & $0.21$ & $1.6$ \\
\hline
  \end{tabular}
  \label{tab:models}
\end{table}

\section{Results}
\label{sec:results}
Table~\ref{tab:models} shows the total number of photons used and the resulting escape fractions. In the following, we will present and discuss the results from each model individually.
\begin{figure*}
\centering
\includegraphics[width=\linewidth]{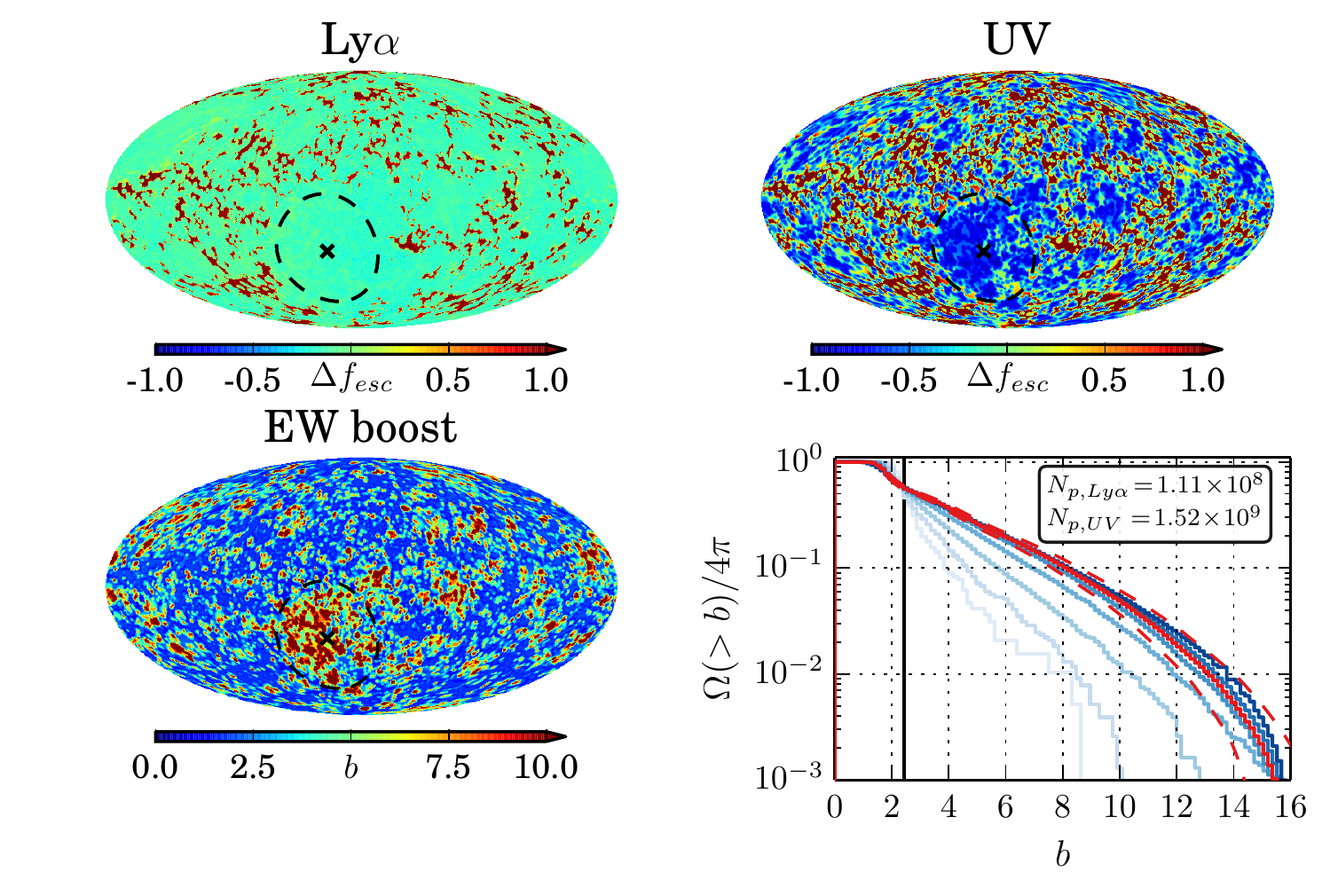}
\caption{Results from the fiducial model. We show  the deviation from uniform escape, $\Delta f_{\rm esc}$ (defined in Eq.~\ref{eq:Deltafesc}), for Ly$\alpha$ (\textit{top left panel}) and for the (non-ionizing) UV-continuum photons (\textit{top right panel}). The black cross and the black dashed line denote the centre and contour of the closest cloud, respectively. \textit{Lower left panel:} Map of the boost factor $b$. The nearest cloud is marked as described above. \textit{Lower right panel:} Cumulative distribution for $b$, i.e. the fraction of the sky that contains a EW-boost $>b$. From light to dark blue the \texttt{HEALPix} parameter $n_{\text{sides}}$ is varied from $4$ to $512$ which corresponds to $\sim(14^2, 28^2, 55^2, 111^2,222^2,443^2,887^2,1774^2)$ directional bins. The red lines mark the binning used for the spherical maps, $n_{\text{sides}} = 128$ (solid) and its average standard deviation for a given $[b, b+{\rm d}b]$ interval (dashed). Additionally, the vertical, black line denotes the photon-weighted $\bar b$.}
\label{fig:fiducial_mutliplot}
\end{figure*} 

\subsection{The fiducial model}
\label{sec:fiducial-model-results}
Our fiducial model yields escape fractions of $0.925$ and $0.381$ for the \Lya and UV-photons, respectively. This results in a photon averaged boost factor of $\bar b = \bar f_{{\rm esc}, Ly\alpha} / \bar f_{{\rm esc}, UV}\approx 2.43$ which is slightly higher than the value obtained by L13 ($\sim 2$). This discrepancy can be explained through geometrical differences -- L13 took the mean of different realizations whereas we used merely one\footnote{L13 noticed significant variations between different realizations (Laursen, private communication)}. Also, the application of an acceleration scheme by L13 could lead to slight variations as discussed above. However, because we study fluctuations around $\bar b$ its numerical value is not very important.

We show our main results in Fig.~\ref{fig:fiducial_mutliplot}. The \textit{upper panels} show the number of \Lya ({\it left panel}) and UV-photons ({\it right panel}) escaping in a given direction. More specifically, we show the deviation from uniform escape, $\Delta f$, which is defined as
\begin{equation}
\Delta f_{\rm esc}(\theta, \phi) \equiv \frac{n_p(\theta, \phi)}{\bar{n}_{p}} - 1,
\label{eq:Deltafesc}
\end{equation} where $n_p(\theta,\phi)$ is the number of photons received in a given directional bin, and $\bar{n}_{p}\equiv \bar f_{\rm esc} N_p / N_{\rm bins}$ in which $N_p$ the total number of photons emitted, and $\bar f_{\rm esc}$ the photon averaged escape fraction. This definition allows the use of the same colour coding while keeping the proportionality to $n_p(\theta, \phi)$ and $N_{p}$. 

The Mollweide projections in Fig.~\ref{fig:fiducial_mutliplot} are obtained using the \texttt{HEALPix} parameter $n_{\rm sides}=128$. This means that we have $N_{\rm bins}=12 n_{\rm sides}^2 = 196608$ equiareal directional bins. This choice leaves the number of \Lya\!\!-photons per bin in the range $[301, 1212]$ (upper left panel) whereas the UV map varies from $136$ to $5658$ photons per bin (upper right panel).  The five, fifty and ninety-five percent quantiles are $(370, 416, 936)$ for the \Lya photons and $(372, 1473, 5339)$ for the UV photons, which shows quantitatively that Ly$\alpha$ escapes more homogeneously than the UV-continuum. In terms of $\Delta f_{\rm esc}$, the quantiles are $(-0.22, -0.12, 0.96)$ for Ly$\alpha$ and $(-0.81, -0.25, 1.7)$ for the UV-continuum.\\

The \textit{lower left panel} of Fig.~\ref{fig:fiducial_mutliplot} shows the spherical distribution of the EW boost factor $b(\theta,\phi) = n_{p,Ly\alpha}(\theta, \phi) N_{p, UV} / (n_{p,UV}(\theta, \phi) N_{p, Ly\alpha})$. Clearly, areas with a relatively high escape fraction for \Lya photons ($\Delta f_{{\rm esc},Ly\alpha}\sim 0$) and a low one for their UV counterparts ($\Delta f_{{\rm esc}, UV}\lesssim -0.5$) results in a high boost factor of $b\gtrsim 5$ (recall that $\bar{b}=2.4$). 
On the other hand, areas with high values of $\Delta f_{{\rm esc},Ly\alpha}$ yield mostly a moderate boost of $b\sim 2$ since in these directions often also $\Delta f_{{\rm esc},UV}$ is large.

The most prominent area with a large boost is clearly due to the cloud located closest to the source. In the projection maps of Fig.~\ref{fig:fiducial_mutliplot}, its centre and contour are marked with a black cross and black dashed line, respectively. This cloud is located in the direction $(\theta, \phi)_{\rm cloud} \approx  (2.10, 0.45)$ at a distance of $\sim 152\,$pc. 

The histogram in the \textit{lower right panel} of Fig.~\ref{fig:fiducial_mutliplot} shows the exact $b$ occurrences. We plot the cumulative fraction of sight lines that have a boost greater than $b$. Different lines represent different angular resolution in the sky. The curves evolve as we increase resolution: this quantifies that increasing angular resolution gives rise to larger fluctuations in individual pixels. This is partly because of the reduced number of photons in each pixel, which increases Poisson noise, and partly physical. In the Appendix we show a Figure which shows the same curves, in case where they are determined completely by Poisson fluctuations (Fig.~\ref{fig:empty_minimulti}). These plots provide a quick check that our results are statistically robust.

Specifically, from light blue to dark blue, the $n_{\rm sides}$ parameter is increased in powers of two from $4$ to $512$. This corresponds to $N_{\rm bins}\approx(192, 768, 3\times 10^3, 1.2\times 10^4, 4.9\times 10^4, 1.97\times 10^5, 7.86\times 10^5, 3.15\times 10^6)$. The red curve highlights the $n_{\rm sides}=128$ or $N_{\rm bins} = 1.97\times 10^5$ case which is used in the projection maps. For this curve we show also the average Poisson uncertainty of all directional bins within a given $[b, b+\mathrm{d}b]$. We give a more detailed explanation of the error estimation in the Appendix.

In addition, a black vertical line indicates the photon-weighted $\bar b = \bar f_{{\rm esc}, Ly\alpha} / \bar f_{{\rm esc}, UV}$ which is equivalent to $N_{\rm bins} = 1$. Note, that the $n_{\rm sides}=512$ and $n_{\rm sides}=256$ are basically identical from the red curve which we interpret as signal of convergence. We tested this by using only a tenth of the available data which resulted in indistinguishably curves for $n_{\rm sides}=4$ to $n_{\rm sides}=256$. The highest binned set ($n_{\rm sides}=512$) possessed more fluctuations due to a lower signal-to-noise ratio.

Focusing on the red curve in the histogram of Fig.~\ref{fig:fiducial_mutliplot}, it is clear that $\sim 10\%$ of the sight lines have a boost factor of $b\gtrsim 8$. A small percentage of them even reaches $b\gtrsim 14$. Recalling the photon-weighted average $\bar b\approx 2.4$, these large values of $b$ are surprising given the spherically symmetric setup of our problem.

\subsection{The realistic model}
\label{sec:realistic-model}
\begin{figure*}
  \centering
  \includegraphics[width=\linewidth]{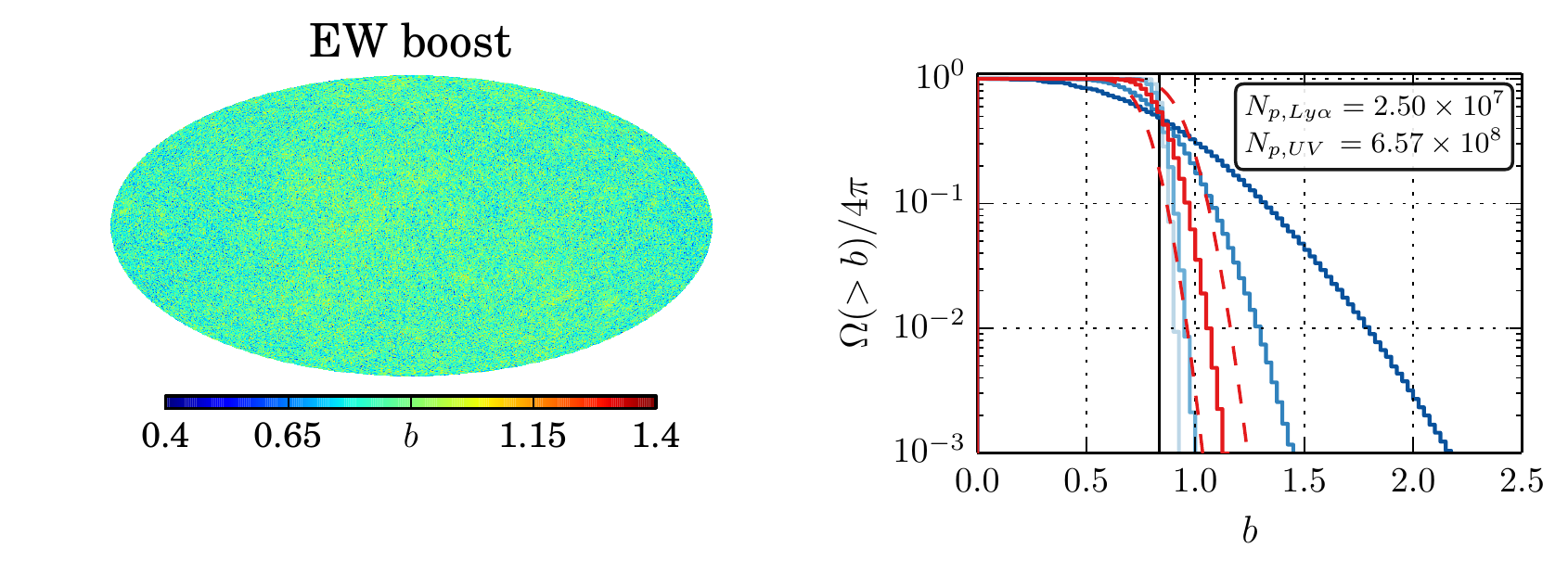}
  \caption{Results from the realistic model. \textit{Left:} Directional dependence of the boost factor $b$. \textit{Right:} Cumulative distribution function of $b$. From light to dark blue the \texttt{HEALPix} parameter $n_{\text{sides}}$ is varied from $32$ to $512$ ($\sim(111^2,222^2,443^2,887^2,1774^2)$ directional bins). The red solid line marks the binning used for the spherical maps, $n_{\text{sides}} = 128$. Again, the red dashed lines mark the $b\pm\langle\sigma_b\rangle$ for a given $[b,b+{\rm d}b]$ range. Additionally, the vertical, black line denotes the photon-weighted $\bar b$.}
  \label{fig:realistic_minimulti}
\end{figure*}
For this model we used $\sim 2.5\times 10^7$ \Lya photons and $\sim 6.5\times 10^8$ UV photons. The photon-weighted escape fractions for the two species are $\bar f_{{\rm esc},Ly\alpha}\approx 0.74$ and $\bar f_{{\rm esc},UV}\approx 0.89$ which yields a boost factor of $\bar b\approx 0.84$. This is in agreement with L13 where most of the realistic runs are in $0.6 \lesssim \bar b \lesssim 1.1$.

Fig.~\ref{fig:realistic_minimulti} summarizes our results: the left panel shows the directional distribution of the EW boost, and the right panel displays the cumulative histogram of the fraction of sight lines greater than $b$. For this plot we varied the \texttt{HEALPix} parameter $n_{\rm sides}$ from $32$ to $512$ resulting in approximately $(1.2\times 10^4, 4.9\times 10^4, 1.97\times 10^5, 7.86\times 10^5, 3.15\times 10^6)$ directional bins. Again, the $n_{\rm sides}=128$ curve is highlighted with red. Also as before, the red dashed lines mark the mean standard deviation per $b$-bin, $\sigma_{\rm b}$.

From these plots, it is clear that the EW boost is much more isotropic in the realistic model than in the fiducial model. In fact, the $b$-distribution is indistinguishable from pure Poisson noise around $\bar b$. For example, the probability $P(b > \bar b + \sigma_b) = P(b - \sigma_b > \bar b) \sim 0.16$ which is what would be expected for pure Gaussian noise\footnote{Another way to see this is by looking at the $(0.05, 0.5, 0.95)$ percentiles, which are $(78, 94, 111)$ for \Lya and $(2860, 2959, 3058)$ for the UV-case continuum. For comparison the same quartiles for an isotropic distribution with Gaussian noise are  $(78, 94, 110)$ for \Lya and $(2870, 2959, 3048)$ for the UV-case continuum, which is practically identical.}. Moreover, we found that the $\Omega(>b)$ distribution emerging from an empty simulation box - from which directional dependent fluctuations arise entirely because of Poisson noise- is close to identical in shape (this distribution is shown in Appendix \ref{sec:empty_sim_box}). 

Another striking feature is the lack of convergence of the curves displayed in the right panel of Fig.~\ref{fig:realistic_minimulti}. This reflects the lack of real anisotropic features, and that these curves are Poisson-noise-dominated: increasing the number of directional bins leads to few photons per bin, and hence to larger uncertainties on the value of $b$. We find the same lack of convergence in the empty test case shown in Appendix \ref{sec:empty_sim_box}).

\begin{figure*}
  \centering
  \includegraphics[width=\linewidth]{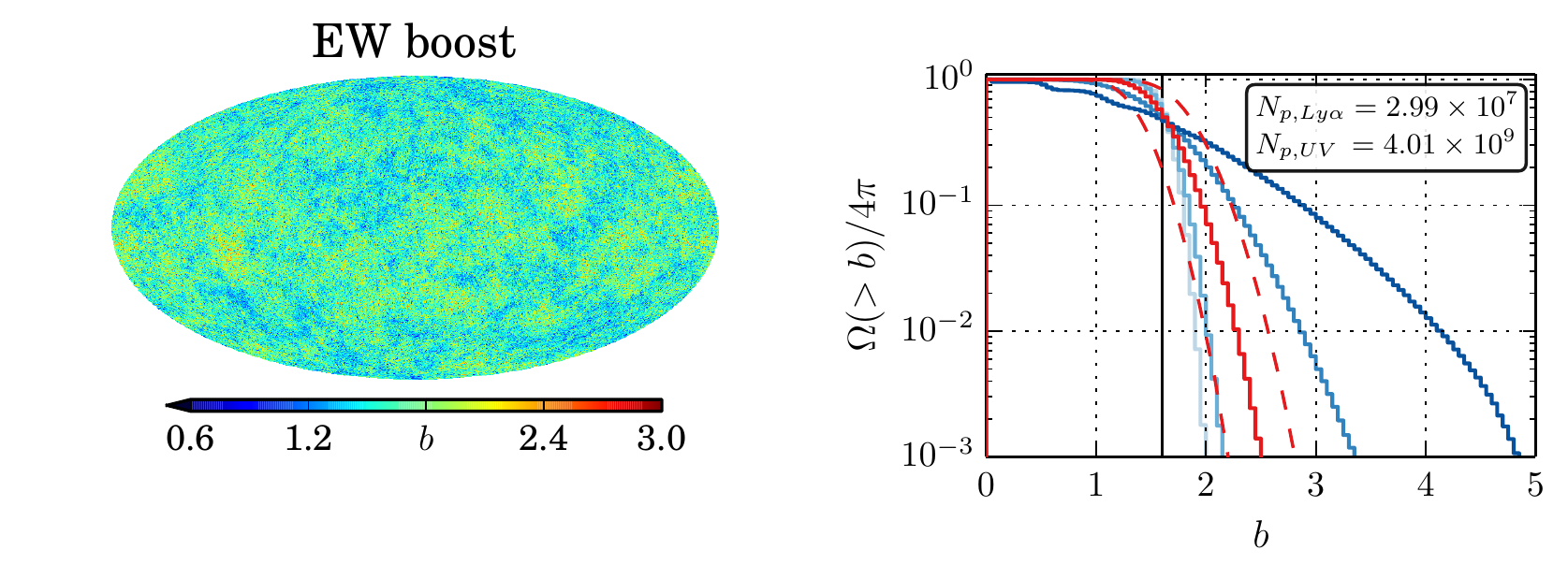}
  \caption{Same as Figure~\ref{sec:realistic-model} but for the `extreme model' (see text).}
  \label{fig:extreme_minimulti}
\end{figure*}

\subsection{The extreme model}
\label{sec:extreme-model}

For the extreme model we use $\sim 30$ million \Lya and $\sim 4$ billion UV-continuum photons. Of those, $\sim 21\%$ UV and $\sim 33\%$ \Lya photons escaped the simulation box. This yields a average boost factor of $\bar{b} \approx 1.60$ which is consistent with the prediction of L13.

We present the results in Fig.~\ref{fig:extreme_minimulti} as before. At first glance, the existence of patches in the projection map suggests that the EW boost in this case is neither as anisotropic as in the fiducial model, nor as isotropic as in the realistic model.  More quantitatively, the $(0.05, 0.5, 0.95)$ quantiles are $(30, 50, 63)$ for \Lya and $(3588,4145,5158)$ for the UV-continuum.  For comparison the same quartiles for an isotropic distribution with Gaussian noise are  $(39, 50, 61)$ for \Lya and $(4041, 4145, 4249)$ for the UV-case continuum. We conclude that directional variations exist, but these are much weaker than in the fiducial model ($1\%$ of the sight lines possess a boost factor $> 2\bar b$ compared to $10\%$ of the sight lines have $\gtrsim 3\bar b$ in the fiducial case). Also, we also do not find convergence when we increase the number of bins, in contrast to the fiducial model. 

\section{Discussion}
\label{sec:discussion}

We investigated the directional dependence of the EW boost factor $b$ in three models labelled as fiducial, realistic and extreme. The fiducial contains strong anisotropies in the emerging EW boost factor. The other two yield a more homogeneous $b$-distribution. Moreover, in the fiducial case the $b$-distribution converges as we increase the angular resolution (see Fig.~\ref{fig:fiducial_mutliplot}). We found over $10\%$ of sight lines with a boost factor which is three times larger than the averaged value of $\bar b \approx 2.4$. In the realistic model this convergence was absent, and consistent with Poisson fluctuations on a uniform distribution. The extreme model yields  an almost-isotropic $b$-distribution (although here the averaged EW boost $\bar b > 1$ by construction). \\

We can understand these results as follows. Several previous studies \citep[e.g.][]{Zheng2013,Zheng2010} have shown that the dominating parameter in various \Lya radiative transfer problems is the hydrogen column density, $N_{\HI}$, along a certain line of sight. We therefore compute this quantity for each directional bin for $n_{\rm sides}=128$ ($N_{\rm bins}=12 n_{\rm sides}^2\approx 2\times 10^5$) and $10$ million randomly drawn sight lines. 
These sight lines start at the photons' emission sites, i.e., in the fiducial case from the centre of the simulation box and in the other two models at a position drawn the spatial distribution (see Sec.~\ref{sec:method}). For the latter two models a single directional bin would on average sample $\sim 10^7/(2\times 10^5) = 50$ emission locations within the simulation box in the absence of scattering. Throughout our discussion, we convert the hydrogen column density $N_{\rm HI}$ to the dust column density or its optical depth. We choose to use the optical depth of the absorbing dust $\tau_a$ because this parameter controls directly the absorption probability and has, hence, a strong impact on the boost value. There exists a one-to-one relation between $\tau_a$ and $N_{\HI}$.\\

\begin{figure}
  \centering
  \includegraphics[width=1.0\linewidth]{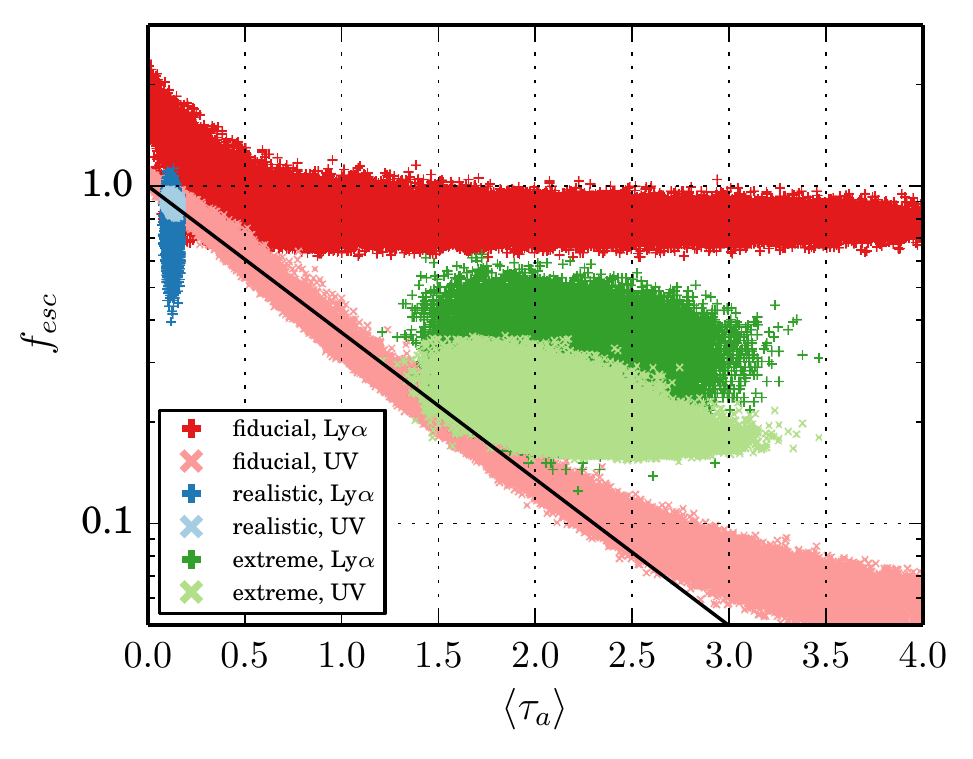}
  \caption{Directional escape fraction $f_{\rm esc} = n_p / (N_p / N_{\rm bins})$ versus the average optical depth in this direction $\langle \tau_a \rangle$. The red scatter points show the fiducial, the blue mark the realistic and green denote the extreme model. Also, the ``plus'' (in darker colours) and ``x'' (lighter colours) symbols display the escape fraction of the \Lya and UV radiation in this direction, respectively. The black line highlights the $\exp(-\langle \tau_a \rangle )$ line.}
  \label{fig:fesc_vs_tau}
\end{figure}

Each directional bin has therefore one associated $\tau_a$ (for the fiducial model) or $\langle \tau_a \rangle$ (for the realistic and extreme models). We compare this quantity to the directional escape fraction $f_{\rm esc}(\theta,\phi) = n_p / (N_p / N_{\rm bins})$ for each bin (see \S~\ref{sec:fiducial-model-results} for a definition of these quantities). We show this comparison in Fig.~\ref{fig:fesc_vs_tau} where each point represents one direction $(\theta,\phi)$. In the fiducial model, $\langle \tau_a\rangle = \tau_a$, because all photons are emitted in the same location. However, in the realistic and extreme models we have $\langle\tau_a\rangle$ is the mean of (one average) $51$ sight lines. \\

The fiducial model data (red in Fig.~\ref{fig:fesc_vs_tau}) presents the Neufeld mechanism in an exemplary way: 
while the escape fraction of the UV photons follows closely the theoretically expected value of $\exp(-\tau_a)$ (solid black line), the \Lya escape rate is unaffected by the directional optical depth and instead constant at $f_{{\rm esc},Ly\alpha}\approx \bar f_{{\rm esc},Ly\alpha}$. In other words, the \Lya photons do not penetrate as deeply into the hydrogen clouds lying on the direct sight line. Instead, their path mainly traces the low-density ICM, which results in a higher escape fraction {\it independent of the direction in which these photons were first emitted}. 
Consequently, sight lines with high $\tau_a$ values possess a large EW boost as the difference between $f_{{\rm esc},Ly\alpha}$ and $f_{{\rm esc}, UV}$ in Fig.~\ref{fig:fesc_vs_tau} shows\footnote{Two departures from this simple correlation are for ({\it i}) large column densities ($\tau_a \gtrsim 2.5$) UV photons escape more efficiently, which arise because UV-continuum photons actually do scatter and can consequently escape in these high-$\tau_{\rm a}$ directions; and ({\it ii}) for very small optical depths ($\tau_a\lesssim 0.3$) the \Lya photons are affected by a similar enhanced $f_{\rm esc}$. The reason for the former comes from the characteristic `weighting' of lower-density directions after a scattering event. This `weighting' of trajectories will be explained in detail below. 
The rise of $f_{{\rm esc},Ly\alpha}$ is also due to photons that are scattered \textit{into} the virtually empty direction -- again with the statistically favouring of these sight lines.}. \\

The sight lines of the realistic and extreme model cover a much smaller range of $\langle \tau_a \rangle $ in Fig.~\ref{fig:fesc_vs_tau} (shown in blue and green, respectively). This is because each $\langle \tau_a \rangle$ represents the mean of on average $N_{\text{sight~lines}}/N_{\rm bins}\approx 51$ sight lines. While in the fiducial case each sight line starts at the identical position, for a spatially extended source one particular direction samples $\sim 51$ randomly selected emission sites throughout the cloud, each of which contain their own $\tau_a$. 

This difference introduces two additional distinctions: \textit{(i)} The scatter of $\langle \tau_a\rangle$ for a given $f_{\rm esc}$ is much larger with an extended source: the $1-\sigma$ dispersion on $\langle \tau_a \rangle$ is only $0.09$ in the fiducial data compared to $0.48$ and $0.57$ in the realistic and extreme data, respectively. \textit{(ii)} The resulting $f_{{\rm esc}, UV}$ depends very weakly on $\langle \tau_a \rangle$. This is most evident for the extreme model (green in Fig.~\ref{fig:fesc_vs_tau}), as the $\langle \tau_a \rangle$-range is too small for the realistic model (shown in blue).  The extreme model typically fulfills $f_{{\rm esc}, UV}\gtrsim \exp(-\tau_a)$, as $f_{\rm esc,UV}$ in a certain direction (and hence a given $\langle \tau_a \rangle$) is weighted towards\footnote{A mathematical way of phrasing this is via the `triangle inequality' which states that $\langle \exp (-\tau_a) \rangle > \exp( - \langle \tau_a \rangle)$ \citep[also see][]{Hansen2005}.} sight lines that pass though low $N_{\HI}$ (we refer to this as `statistical favouring'). That is, UV photons emerging the simulation box in a certain direction are more likely to originate from a location with $\tau_a\lesssim\langle\tau_a\rangle$. This characteristic favouring of lower density trajectories plays a key role in the disappearance of the $b$ anisotropy when introducing an extended source. 

The \Lya data of the realistic and extreme are ordered in a fundamentally different way. In the extreme model, both the Neufeld effect and statistical favouring cause the data points (shown in dark green in Fig.~\ref{fig:fesc_vs_tau}) to lie mainly above the $\exp(-\langle\tau_a\rangle)$ line. In the realistic model (shown in dark blue), $f_{{\rm esc},Ly\alpha}$ stretches far below this line. This further shows that the Neufeld mechanism is not active in this case. That is, the \Lya photons' trajectories reach far into the clouds and a long path is required for eventual escape.\\

In summary, for spherical distributions of dusty clumps two requirements have to be fulfilled in order to give rise to an anisotropic boost factor:
\begin{enumerate}
\item The photon emitting source may not be too extended. Otherwise, anisotropies that may exist in the UV luminosity -- and as a consequence the EW boost -- are washed out.
\item The so called ``Neufeld mechanism'' has to be active. Without it, the directional distribution of the \Lya photons follows that of UV-continuum too closely, and no sight line possesses an exceptional boost factor.
\end{enumerate}
In order to verify these two requirements, we first ran the fiducial model with an extended source and several values for $H_{\rm em}$. Doing this, we find that already $100\,{\rm pc}< H_{\rm em}< 500\,$pc is sufficient to wipe out all anisotropies existent for smaller scale lengths. We expect the limiting value to be roughly the inter-cloud spacing which is $\sim 330\,$pc in this case.

Secondly, we reversed the setup and simulated a point source in the realistic and extreme models. As a result, the averaged boosts $\bar b$ increase to $\bar{b} \sim 1.0$ and $\bar{b}\sim 4.4$, respectively. Also, as expected, the $b$ distribution of the modified extreme model yields big anisotropies with $\Omega(>3\bar b)/4\pi\approx 0.1$ whereas its counterpart of the modified realistic model remains isotropic.

\section{Conclusions \& Outlook}
\label{sec:conclusion}

We have explored the directional dependence of the escape of Ly$\alpha$ and (non-ionizing) UV-continuum from a `multi-phase' medium, and have quantified what fluctuations in the Ly$\alpha$ equivalent width (EW) this may introduce. The goal of this analysis was to address whether directional fluctuations in Ly$\alpha$ EW can lead to substantial departures from the photon-averaged value, and whether this may help to explain `unusually' large observed values of the EW.

Our multiphase medium consisted of spherical distributions of dusty clumps within a hot dust-free medium. Our models are taken from a recent analysis by \citet{Laursen2012}, who provide a detailed analysis of what are physically reasonable ranges of the parameters describing such a clumpy medium. We focus on three models: ({\it i}) the `fiducial' model corresponds to the fiducial model in \citet{Laursen2012}, which facilitates comparison with their previous work; ({\it ii}) the `realistic' model corresponds to a model which contains physically reasonable parameters for the multiphase gas, and which have an angle-averaged EW boost  $\langle b \rangle < 1$; ({\it iii}) the `extreme' model corresponds to one of the few models in \citet{Laursen2012} in which the Neufeld-mechanism is active, i.e. $\bar b > 1$.

We performed Ly$\alpha$ Monte-Carlo radiative transfer calculations with $\sim 10^7-10^8$ Ly$\alpha$ photons -- with the standard acceleration schemes turned off -- and $\sim 10^8-10^9$ UV-continuum photons for these three models\footnote{We performed these calculations on $\sim 200$ cores.}. We stored the propagation directions of each Ly$\alpha$ and UV-photon as they escape from the multiphase gas, and use this information to make $2$D EW-boost maps (shown in Fig.~\ref{fig:fiducial_mutliplot}--~\ref{fig:extreme_minimulti}).

We found that directional variations were very large in the fiducial model, with $\sim 10\%$ [$\sim 1\%$] of the sight lines having a boost $b > 3\bar{b} \sim 7$ [$b > 5\bar{b}\sim 12$], while the average value was only $\bar b \sim 2.4$. However, in both extreme and realistic models the fluctuations were significantly reduced, and especially in the realistic model the fluctuations were consistent with Poisson noise. The main reason for this difference is the spatial extend of the Ly$\alpha$ and UV-source in these models: the fiducial model contained a central point source, while the other two models contained sources that were spatially extended. This has major implications: we showed that in the fiducial model the nearest dusty clump effectively blocked UV-continuum photons from escaping. This clump thus casts a UV-continuum `shadow' on the sky. In contrast, we found that Ly$\alpha$ photons escaped (much) more isotropically, which therefore enhanced the Ly$\alpha$ EW in directions corresponding to the UV-continuum shadow. Each point source within a simulation induces a similar pattern of UV-continuum shadows on the sky. When averaged over a large number of sources, the directionally dependence of the EW is reduced. We found that the EW-anisotropies vanish rapidly when the spatial extend of the Ly$\alpha$ and UV-continuum\footnote{Since elevated EW regions correspond to regions of suppressed UV-continuum escape, it is really the spatial extend of the UV-continuum source that is relevant: to boost the EW in regions of suppressed UV-continuum escape we need isotropic escape of Ly$\alpha$ photons, which is more easily achieved for a spatially extended Ly$\alpha$ source. In this work however, we have assumed for simplicity that the Ly$\alpha$ and UV-continuum source have the same spatial extent.} source starts to exceed the mean clump separation. This presents a key quantity determining in the directional-dependence of the EW.

Our results can be translated into a physical picture that directionally-dependent EW boosting can be important in cases where the source that dominates the UV-continuum luminosity of a galaxy is obscured from us by an \HI cloud that contains dust. In such a scenario, the dusty cloud suppresses the observed UV-continuum flux. Scattering of \Lya photons off \HI atoms in the ISM can cause Ly$\alpha$ photons to escape from the galaxy more isotropically. Large directional EW boosts would likely still require an efficient (isotropic) escape of Ly$\alpha$ photons, and would thus benefit from having a multiphase ISM. However, this multiphase ISM would not have to produce $\bar{b}>1$ -- which previous studies found to be difficult to achieve physically\footnote{Perhaps an even simpler way to illustrate this point is that we can take a model with parameters that L13 claims to be reasonable. This model would typically produce $\bar{b}<1$. If we add one single clump with an enhanced dust content to that model such that it covers a significant fraction of the UV source, then this cloud can cast a UV-shadow and boost the EW in these directions.} -- to give rise to elevated EW values in certain directions. In future work, we will explore this directional EW-boosting in more realistic models of the multiphase ISM, starting with non-spherical clump distributions and clump distributions that have a covering factor $f_{\rm c} <1$, which appears favoured by observations of low-ionisation UV metal absorption lines \citep[e.g.][]{Heckman11,Jones13}.
Anisotropic EW boosting should observationally be very similar to isotropic EW boosting in the sense that large EWs correspond to low values for $f_{\rm esc,UV}$, and therefore the amount of dust extinction/reddening.
We defer a detailed study of observational signatures of directional dependent EW-boosting to this future work.

\section*{Acknowledgments}
We thank Peter Laursen for helpful correspondence, and for providing details about the work in L13.
We also thank the reviewer for the prompt and constructive feedback.
Some of the results in this paper have been derived using the \texttt{HEALPix} \citep{Gorski2005a} package.

\bibliography{references}

\appendix
\section{Empty simulation box}
\label{sec:empty_sim_box}
\begin{figure}
  \centering
  \includegraphics[width=\linewidth]{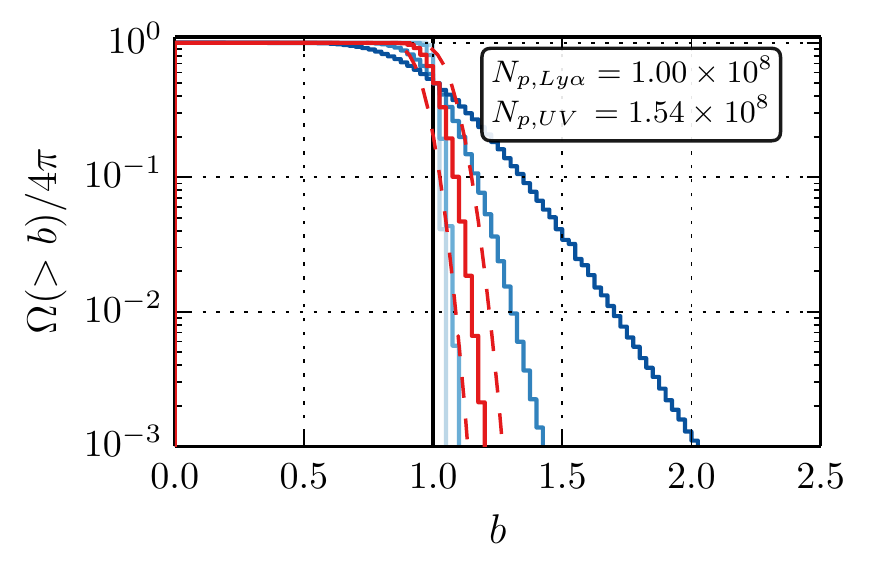}
  \caption{The cumulative histogram displayed also in Fig.~\ref{fig:fiducial_mutliplot}--\ref{fig:extreme_minimulti}, however, for an empty simulation box. }
  \label{fig:empty_minimulti}
\end{figure}
In order to test the code and the statistical significance of our findings, we repeated our analysis on an empty simulation box. In this model, Ly$\alpha$ and UV-continuum photons escape isotropically. However, the number of Ly$\alpha$ photons and UV-continuum photons that strike a certain bin on the sky fluctuates due to the finite number of photons utilised in the simulation. For example, for $N_{p,{Ly}\alpha}=10^8$ Ly$\alpha$ photons and $n_{\rm sides}=128$ we expect ab average of $\bar n_{{\rm Ly}\alpha}=10^8/(12\times 128^2)\approx 509$ photons per bin and a boost average $\bar b = \bar f_{{\rm esc},Ly\alpha}/\bar f_{{\rm esc},UV}=1$. The variance on the boost $b$ can be estimated from
\begin{align}
\sigma^2_{\rm b}&= \Big{(} \frac{\partial b}{\partial f_{\rm esc,UV}}\Big{)}^2\sigma^2_{f_{\rm esc,UV}}+ \Big{(} \frac{\partial b}{\partial f_{{\rm esc},Ly\alpha}}\Big{)}^2\sigma^2_{f_{{\rm esc},Ly\alpha}}\\
&= b^2\left(\frac{1}{n_{Ly\alpha}} + \frac{1}{n_{UV}}\right),
\end{align} where $f_{{\rm esc},UV}$ and $n_{UV}$ denote the escape fraction and number of photons of UV-continuum photons in one direction, respectively. The escape fraction is given by $f_{{\rm esc},UV}=n_{UV}/(N_{p,UV}/N_{\rm bins})$ with $N_{p,UV}$ being the total number of UV photons emitted. Naturally, $f_{{\rm esc},Ly\alpha}$ and $n_{Ly\alpha}$ have the same meaning for Ly$\alpha$-photons.

Because $n_{Ly\alpha}$ and $n_{UV}$ are random Poisson variables and $b$ is a ratio of both, the probability density function of $b$ is more complicated. However, when assuming a large number of photons per bin ($\gtrsim 100$) the two probability density functions converge to Gaussians and an analytic estimate of the resulting $b$-distribution can be found.
The resulting distribution is -- to leading order -- again a Gaussian\footnote{More quantitatively, the distribution of $b$ in an empty box can be written as $b\approx\bar b \frac{1+X_{Ly\alpha}/\bar n_{Ly\alpha}}{1+X_{UV}/\bar n_{UV}}\approx (1+X_{Ly\alpha}/\bar n_{Ly\alpha}-X_{UV}/\bar n_{UV})$ where $X \equiv n-\bar{n}$ denote normal distributed variables with zero mean, and a variance of $\bar{n}$. We further used that $X \ll \bar{n}$. The parameter $b$ can thus be expressed as the sum of the two Gaussian variables, and is thus Gaussian itself with its variance being the sum of the two variances (i.e. $\bar{n}^{-1}_{UV}+\bar{n}^{-1}_{{\rm Ly}\alpha}$).} centreed at $\bar{b}=1$ with variance $\sigma_{\bar b}^2=\bar b^2(\bar n_{Ly\alpha}^{-1}+\bar n_{UV}^{-1})$. For the above mentioned $\bar n_{Ly\alpha}\approx 509$ this yields $\sigma_{\bar b}\approx 0.057$ which agrees very well with our numerical results for the empty box.

These considerations allow us to quantitatively understand the plot in Fig.~\ref{fig:empty_minimulti} in the same fashion as used in Fig.~\ref{fig:fiducial_mutliplot}--\ref{fig:extreme_minimulti}. For instance, we expect $16\%$ [$2.5\%$] of sight lines to lie $\sigma_{\rm \bar b}$ [$2\sigma_{\rm \bar b}$] above the average. As we increase $n_{\rm sides}$, we increase $\sigma_{\rm \bar b}$ and our plots `widen'. Here, the rise in $\Omega(>b)$ with increasing number of bins is merely due to the Poisson noise. Figure~\ref{fig:empty_minimulti} nicely illustrates that the curves do not converge as we increase the angular resolution.

\label{lastpage}
\end{document}

%% file: ads_commands.tex
\newcommand\apj{\rmfamily{ApJ}}%
\newcommand\apjl{\rmfamily{ApJ}}%
\newcommand\aap{\rmfamily{A\&A}}%
\newcommand\mnras{\rmfamily{MNRAS}}%
\newcommand\nat{\rmfamily{Nature}}%